\documentclass[FBSedit,FBSmath,ecsub]{FBSsuppl}
\usepackage{amsmath}
\usepackage{amsfonts}
\usepackage{amssymb}

\title{Relativistic Study of Mesonic Baryon Resonance
Decays}
\author{T. Melde, R. F. Wagenbrunn, W. Plessas}
\institute{
Institut f\"ur Theoretische Physik,
Universit\"at Graz, Universit\"atsplatz 5,
A-8010 Graz, Austria
}

\begin{document}
\maketitle
\begin{abstract}
Mesonic baryon resonance decays are calclulated from constituent
quark models along a Poincar\'e-invariant generalization of the
elementary emission model. Covariant results of pionic decay
widths are presented for the Goldstone-boson-exchange constituent
quark model.
\end{abstract}
\section{Introduction}
Investigations of mesonic resonance decays have a long tradition,
and in the focus of interest have been, notably, the performances
of various constituent quark models (CQMs) as well as the adequacy
of different decay operators for the
mechanism of meson creation/emission. Despite considerable
efforts invested one has still not yet arrived at a satisfactory
explanation especially of the $N^*$ and $\Delta$ resonance decays.
Also complementary attempts beyond the CQM approach have not
succeeded much better with hadron decays, and
more generally, with providing a comprehensive working model
of low-energy hadronic physics based on QCD.
This situation is rather disappointing from the theoretical side,
especially in view of the large amount of experimental data
accumulated over the past years and the ongoing high-quality
measurements at such facilities as JLAB, MAMI and others (for a an
overview of the modern developments see the proceedings of the
past $N^*$ Workshop~\cite{Drechsel:2001}).

In a recent CQM the interaction between two constituent quarks was
based on Goldstone-boson exchange (GBE)~\cite{Glozman:1998ag} in
addition to a linear confinement. The so-called GBE CQM at the same
time fulfills the requirements of Poincar\'e invariance and thus it paves
the way to a relativistic treatment of reactions involving baryons.
Specifically, due to the usage of a relativistic kinetic-energy operator, the
Hamiltonian of the GBE CQM leads to an invariant mass operator in
relativistic quantum mechanics, where the quark-quark interactions are
introduced via the Bakamjian-Thomas construction~\cite{Keister:1991sb}.

Up till now the GBE CQM has already been put to some tests in calculating
mesonic decays of resonances of light and strange baryons in a
non-relativistic
framework~\cite{Krassnigg:1999ky,Plessas:1999nb,Theussl:2000s}.
Here, we calculate the mesonic decay widths in a covariant approach
along a generalization of the elementary emission model (EEM) in
point-form relativistic quantum mechanics~\cite{Klink:1998pr}.
We compare the predictions of the GBE CQM to
analogous results from a CQM with a one-gluon-exchange (OGE) hyperfine
interaction, namely the relativized Bhaduri-Cohler-Nogami CQM
as parametrized in ref.~\cite{Theussl:2000s}.
\renewcommand{\arraystretch}{1.5}
\begin{table}[b!]
\caption{Predictions for pionic decay widths by the GBE 
CQM~\cite{Glozman:1998ag} along the
EEM in PFSA in comparison to experiment, an analogous calculation with
the OGE CQM~\cite{Theussl:2000s}, and results from a non-relativistic
EEM approach.}
\label{tab1}
\begin{tabular}{|c|r|c|c|c|c|}
\hline
       &      &  \multicolumn{2}{|c|}{Rel. PFSA}
       &  \multicolumn{2}{|c|}{Nonrel. EEM } \\
Decays & Experiment~\cite{pdg} & \multicolumn{2}{|c|}{\small CQM} &
\multicolumn{2}{|c|}{\small GBE CQM} \\
       &  & {\small GBE} & {\small OGE} & {\small dir} & {\small dir+rec}  \\
\hline
{\small $N^{\star}_{1440}\rightarrow \pi N_{939} $}
	&  $\left(227\pm 18\right)_{-59}^{+70}$
	         &  $30.3$
             &  $37.1$
			 &  $4.85$
			 &  $6.16$ \\
\hline
{\small $N^{\star}_{1520}\rightarrow \pi N_{939} $}
	 & { $\left(66\pm 6\right)_{-5}^{+9}$  }
             &  $16.9$
			 &  $16.2$
			 &  $22.0$
			 &  $38.3$
\\
\hline
{\small $N^{\star}_{1535}\rightarrow \pi N_{939} $ }
	 & { $ \left(67\pm 15\right)_{-17}^{+55}$ }
             &  $93.2$
			 &  $122.8$
			 &  $24.3$
			 &  $574.3$
\\
\hline
{\small $N^{\star}_{1650}\rightarrow \pi N_{939}$ }
	 & { $ \left(109\pm 26 \right)_{-3}^{+36}$ }
             &  $28.8$
			 &  $38.3$
			 &  $11.3$
			 &  $160.3$
\\
\hline
{\small $N^{\star}_{1675}\rightarrow \pi N_{939} $}
	 & { $ \left(68\pm 8\right)_{-4}^{+14}$ }
             &  $5.98$
			 &  $6.20$
			 &  $7.65$
			 &  $15.1$
\\
\hline
{\small $N^{\star}_{1700}\rightarrow \pi N_{939} $ }
	 & { $ \left(10\pm 5\right)_{-3}^{+3}$ }
             &  $0.91$
			 &  $1.19$
			 &  $1.43$
			 &  $2.87$
\\
\hline
{\small $N^{\star}_{1710}\rightarrow \pi N_{939} $}
	 & { $\left(15\pm 5\right)_{-5}^{+30}$  }
             &  $4.06$
			 &  $2.28$
			 &  $23.4$
			 &  $5.95$
\\
\hline
{\small $\Delta_{1232}\rightarrow \pi N_{939} $}
	 & { $\left(119\pm 1 \right)_{-5}^{+5}$ }
             &  $33.7$
			 &  $32.1$
			 &  $59.1$
			 &  $81.2$
\\
\hline
{\small $\Delta_{1600}\rightarrow \pi N_{939} $ }
	 & { $\left(61\pm 26\right)_{-10}^{+26} $ }
             &  $0.116$
			 &  $0.503$
			 &  $74.2$
			 &  $55.7$
\\
\hline
{\small $\Delta_{1620}\rightarrow \pi N_{939} $}
	 & { $\left(38\pm 8 \right)_{-6}^{+8}$ }
             &  $10.4$
			 &  $14.6$
			 &  $4.82$
			 &  $74.8$
\\
\hline
{\small $\Delta_{1700}\rightarrow \pi N_{939} $ }
	 & { $ \left(45\pm 15\right)_{-10}^{+20}$ }
             &  $2.92$
			 &  $3.10$
			 &  $7.12$
			 &  $14.4$
\\
\hline
\end{tabular}
\end{table}
\section{Theory}
Generally, the decay width of a particle is defined by the expression
\begin{equation}
	\Gamma=2\pi \rho_{f}\left| F\left(i\rightarrow
	f\right)\right|^{2},
\end{equation}
where $ F\left(i\rightarrow f\right)$ is the transition amplitude
and $\rho_{f}$ is the phase-space factor. In Eq. (1) one has to
average over the initial and to sum over the final
spin-isospin projections.
In non-relativistic calculations of baryon resonance decays one has
usually made an arbitrary choice of the phase-space factor.
In the rest frame of the decaying resonance, either a non-relativistic
form, $\rho_{f}=2\pi \frac{M_{f}M_{\pi}}{M_{i}}q$,
or a relativistic form $\rho_{f}=2\pi \frac{E_{f}E_{\pi}}{M_{i}}q$
has been used. Herein, $M_{i}$ is the mass of the initial state and
$M_{f},M_{\pi}$ as well as $E_{f},E_{\pi}$ are the masses and energies
of the decay products, the nucleon and the pion, respectively.
In the present work we follow a Poincar\'e-invariant description of the
transition amplitude resulting in a unique choice of the phase-space
factor.

For the actual calculation in the point form it is advantageous to
introduce so-called velocity states
\begin{multline}
\left|v;\vec{k}_1,\vec{k}_2,\vec{k}_3;\mu_1,\mu_2,\mu_3\right\rangle=U_{B(v)}
\left|k_1,k_2,k_3;\mu_1,\mu_2,\mu_3\right\rangle=
\\
\prod\limits_{i=1}^3D^{\frac{1}{2}}_{\sigma_i\mu_i}[R_W(k_i,B(v))]
\left|p_1,p_2,p_3;\sigma_1,\sigma_2,\sigma_3\right\rangle,
\end{multline}
where $B\left(v\right)$ is a boost with four-velocity $v$ and
$U_{B\left(v\right)}$ its unitary representation. The boosted momenta
are defined by $p_{i}=B\left(v\right)k_{i}$, where
$k_{i}=\left(\omega_{i},{\bf k}_{i}\right)$ and
$\sum{\vec k_{i}}=0$;
the $D^{\frac{1}{2}}$ are the spin-$\frac{1}{2}$ representation matrices
of Wigner rotations $R_{W}\left(k_{i},B\left(v\right)\right)$. The
baryons are described by simultaneous eigenstates of the 
four-momentum operator (or equivalently the mass operator), the 
total-angular-momentum operator, and its z-component; we denote them by
$\left|P,J,\Sigma\right>$. The transition amplitude is then defined in
a Poincar\'e-invariant fashion, under overall momentum conservation
($P'-P=Q_{\pi}$), by
\begin{eqnarray}
F\left(i\rightarrow f\right)
& = &
\left< P,J,\Sigma\right|
       \hat {\cal D}_{\alpha}
       \left|P',J',\Sigma'\right>
       \nonumber\\
        &\sim & 
\int{      d^{3}k_{2}d^{3}k_{3}d^{3}k'_{2}d^{3}k'_{3}
	     }
	     \nonumber
	     \\
	     & &
	      \Psi^{\star}_{MJ\Sigma}
  \left(
  \vec k_{1},\vec k_{2},\vec k_{3};\mu_{1},\mu_{2},\mu_{3}
  \right)
  \Psi_{M'J'\Sigma'}
  \left(
  \vec k'_{1},\vec k'_{2},\vec k'_{3};\mu'_{1},\mu'_{2},\mu'_{3}
  \right)
	     \nonumber \\
	     & &
	     \prod_{\sigma_{i}}{
	     D^{\frac{1}{2}\star}_{\sigma_{i}\mu_{i}}
	     \left[R_{W}\left(k_{i},B\left(v_{in}\right)\right)\right]}
	     \prod_{\sigma'_{i}}{
	     D^{\frac{1}{2}}_{\sigma'_{i}\mu'_{i}}
	     \left[R_{W}\left(k'_{i},B\left(v_{f}\right)\right)\right]}
	     \nonumber
	     \\
	     & &
	     \left<p_{1},p_{2},p_{3};\sigma_{1},\sigma_{2},\sigma_{3}
	     \right|
	     \hat {\cal D}_{\alpha}
	     \left|p'_{1},p'_{2},p'_{3};\sigma'_{1},\sigma'_{2},\sigma'_{3}
	     \right>,
\end{eqnarray}
where the baryon wave functions $\Psi^{\star}_{MJ\Sigma}$ and
$\Psi_{M'J'\Sigma'}$ enter as
velocity-state representations of the baryon states 
$\left<P,J,\Sigma\right|$ and $\left|P',J',\Sigma'\right>$, 
respectively.
In a first attempt, we investigate a decay operator, which can be
interpreted as a generalization of the EEM.
Namely, we assume that a pion is created on one of the
quarks, while the other two quarks of the decaying baryon resonance
are merely spectators. Consequently, the decay
operator is taken in point-form spectator approximation (PFSA).
Here, we assume a pseudo-vector coupling for pointlike particles.
This yields the decay operator in the form
    \begin{multline}
    \left<p_{1},p_{2},p_{3};\sigma_{1},\sigma_{2},\sigma_{3}\right|
    \hat {\cal D}_{\alpha}
    \left|p'_{1},p'_{2},p'_{3};\sigma'_{1},\sigma'_{2},\sigma'_{3}\right>
    \\
    \sim 3i g_{q\pi}
    \bar u\left({p_{1}},\sigma_{1}\right)
    \gamma^{5}\gamma^{\mu}{\vec \lambda^{F}}
    u\left({ p'_{1}},\sigma'_{1}\right)
    \\
    \times
    2{p'}^{0}_{2} \delta\left(\vec p_{2}-\vec p'_{2}\right)
    2{p'}^{0}_{3}\delta\left(\vec p_{3}-\vec p'_{3}\right)
   \delta_{\sigma_{2}\sigma'_{2}}
   \delta_{\sigma_{3}\sigma'_{3}}Q_{\mu},
   \end{multline}
where $g_{q\pi}$ is the pion-quark coupling constant and $\vec
\lambda^{F}$ the flavour operator.
It should be noted that in PFSA the impulse delivered
to the quark that emits the pion is not equal to the impulse
delivered to the baryon as a whole. The momentum transfer $\tilde q$ to
this single quark is uniquely determined from the momentum $Q$
transferred to the baryon and the two spectator conditions.
\section{Results}
In Table~\ref{tab1} we present the predictions of the GBE CQM for
pionic decay widths of $N^{*}$ and $\Delta$ resonances. They have been
calculated directly from the CQM wave functions with the EEM decay
operator in PFSA; no further parametrization has been introduced.
As is immediately evident, only in a few instances a reasonable
agreement with experiment is found. In most cases the decay widths
remain far from the data, with the theoretical values always being
smaller than the measured ones. These characteristics of the
relativistic decay widths are in contrast to the ones calculated
along the EEM in a non-relativistic approach plus first-order
relativistic (recoil)
corrections~\cite{Krassnigg:1999ky,Plessas:1999nb}.
On the other hand, the OGE and GBE CQMs produce quite similar results
indicating that dynamical effects from the CQM are of lesser
importance than relativistic effects.

In any case, we find the relativistic results to exhibit a completely
new behaviour. This might largely be due to Lorentz boost effects,
which have here been included exactly. Certainly, the predictions for
the pionic decay widths are by no means satisfactory. This hints to
persisting deficiencies in the decay operator and/or the CQM wave
functions. The latter, relying on $\{QQQ\}$ configurations only,
are tentatively not realistic enough especially for the decaying
resonance state. Further work will have to clarify these shortcomings.
\vspace{-2mm}
\begin{acknowledge}
This work was supported by the Austrian Science Fund (Project P14806).
\end{acknowledge}
\end{document}